\documentclass[twocolumn,showpacs,prl,amsmath,amssymb]{revtex4}   
\usepackage{amsmath}
\usepackage{amsfonts}
\usepackage{amssymb}
\usepackage{graphicx}

\begin{document}

\title{The nature and strength of inter-layer binding in graphite}

\author{Leonardo Spanu$^{1}$,  Sandro Sorella$^{2}$,  Giulia Galli$^{1}$}
\affiliation{$^{1}$ Department of Chemistry, University of California at Davis, Davis, CA\\
$^{2}$ International School for Advanced Studies SISSA-ISAS via Beirut 3-5, Trieste Italy}

\begin{abstract}
We  computed the inter-layer bonding properties of graphite using an 
ab-initio many body theory. We carried out variational and diffusion 
quantum Monte Carlo calculations and found an equilibrium inter-layer 
 binding energy in good agreement with most recent experiments. 
We also analyzed the behavior of the total energy as a function of interlayer separation  at 
large distances comparing  the results with  the predictions of the random phase approximation. 
\end{abstract}


\maketitle
% introduction

The excitement generated by the ability to fabricate graphene layers and possibly to tune their electronic properties has renewed great interest in weak interactions in graphitic systems \cite{castro}. The hope of using graphene as a component for next generation electronics relies, among other things, on a detailed understanding and control of how it interacts with its surrounding (e.g. with supporting substrates)\cite{gg2}.

Unfortunately the nature and strength of binding in graphitic materials are poorly understood. 
For example, large uncertainties are associated with a fundamental physical quantity such as the strength of inter-layer binding in graphite. In addition, the way the interaction between planes decays as a function of distance is controversial\cite{dobson}\cite{DobsonPRB2009} , casting doubts on our current understanding  of weak binding in carbon based system. Both binding strength and power law behavior of interlayer interactions in graphite are relevant to the comprehension of a multitude of materials, including graphite intercalation compounds, novel nano-electronic components, and carbon based systems for hydrogen storage.

 From a theoretical standpoint, unravelling binding in graphitic systems is intimately related to understanding the role played by dispersion forces, and to acquiring the ability to describe these purely quantum mechanical interactions at a high level of accuracy. 
Local (LDA)\cite{KS}  or semi-local \cite{gga} approximations to Density Functional Theory (DFT) do not correctly describe  long range correlation, due to the local character of the exchange and correlation potential. Progress ~\cite{langreth} has been recently made in including dispersive interactions within a DFT formalism, in a self consistent, {\it non empirical} manner and binding between graphene layer has been predicted. In  Ref. \cite{langreth} the authors report an interlayer distance overestimated by $7\%$ with respect to experiment, and a binding energy of $45.5$ meV/atom consistent with most extrapolated measurements of exfoliation energy \cite{girifalco}. (Note that the calculations of Ref. \cite{langreth} are for two isolated graphene layers, not for a graphite solid). Semi empirical methods \cite{empirical},\cite{grimme}  have been often used  to treat dispersion correlations where DFT energies are corrected with a contribution coming from attractive $C_6/R^6$ potentials between pair of nuclei. However the power law behavior used in empirical and semi-empirical approaches to describe non retarded dispersion forces  has been recently questioned \cite{dobson}.

 At present, no direct measurement of graphite binding energy is available. However estimates based on theoretical models have been reported in the literature, using experimental data for exfoliation energies (EE, the energy required to remove one graphene plane from the surface of a graphite solid). Three experiments have reported data for EE \cite{zacharia}\cite{benedict}\cite{girifalco}, and a common aspect to the  experimental analyses is the use of simple, fitted  force fields to model either C-C or C-H interactions. 
The early work of Girifalco and Lad \cite{girifalco} gave a EE value of $43(5)$ meV/atom. Using a Lennard-Jones potential, the difference between exfoliation and cleavage energy (that is the interaction between two semi-infinite crystals) was estimated to be $18\%$ but the exact difference remains unknown. 
The work by Benedict {\it et al.} \cite{benedict} extrapolated the interaction energy between graphite layers ($33$ meV/atom) from measurements on collapsed nanotubes, using a force field to model the tubes’ elastic properties.  More recently Zacharia {\it et al.} \cite{zacharia} performed detailed desorption experiments of aromatic molecules from a graphite surface. The graphite EE was derived by extrapolating the molecules’ EE as a function of the number of carbon atoms, thus obtaining a value of $52(5)$ meV/atom. This yields an estimate of the cleavage energy ($62$ meV/atom) which is twice as
much as that reported in Ref. \cite{benedict}. 
  
 Given the state of experiment and theory in determining the binding in graphite, there is a clear need for accurate calculations, eliminating as much as possible all approximations used so far in the literature, and possibly providing guidance to future experiments.  Here we report the binding curve of graphite in AB stacking as obtained using quantum Monte Carlo (QMC) calculations, that is a many-body computational technique \cite{review} capable of accounting for dispersion forces \cite{drummond_needs},\cite{benzene}. We obtain an equilibrium inter-layer bond distance in satisfactory agreement with experiment ( 3.426(36)\AA\  versus 3.35\AA\ ), and a  binding energy \footnote{The BE is defined  as $E_{n-layer} - n*E_{1-layer}$, where $E_{n-layer}$ is the total energy of a graphite periodic cell containing $n$ layers at the optimized, equilibrium position and $E_{1-layer}$ is the total energy of a graphene layer} 
 of $56(5)$ meV/atom, in accord with the mesaurements of Zacharia {\it et al.} \cite{zacharia}.  We find that at distances between $4$ \AA\  and $8$\AA\  the total energy curve exhibits a $ \sim D^{-4.2}$ behavior as a function of inter-layer spacing $D$, which is very similar to that predicted by the Random Phase Approximation (RPA)  applied to two-dimensional (2D) semi-conducting layers\cite{dobson}.

In our investigation we have carried out variational Monte Carlo (VMC) and Lattice Regularized Diffusion Monte Carlo (LRDMC) calculations \cite{casula_lrdmc} with the {\it TurboRVB} code \cite{TRVB}. Our many body wavefunction is the 
product of a  Slater Determinant and a Jastrow many body factor. 
The determinant is obtained with $N/2$   molecular orbitals $\psi_j( \vec r)$, each doubly occupied by opposite 
spin electrons ( $N$ is the total number of electrons).  The orbitals  $\psi_j(\vec r)  $ are expanded  in a Gaussian 
single-particle basis set $\{\phi _{i}\}$, centered on atomic  nuclei,  i.e.
$\psi_j (\vec r)=\sum_{i}\lambda _{i,j}\phi _{i}({r})$

Electron correlation effetcs are  included in our wave function (WF)
through the Jastrow factor      
$J (\vec r_1, \cdots , \vec r_N) 
= \prod\limits_{i<j} \exp ( f( \vec r_i, \vec r_j ) $, where 
$f( \vec r,\vec r^\prime)$ depends only upon  two-electron coordinates.  
The   function $f$ is expanded in a basis of Gaussian atomic orbitals $\bar \phi_i $,:
%\begin{equation}
$f( \vec r, \vec r^\prime) = \sum_{i,j}  g_{i,j} \bar \phi_i (\vec r) \bar \phi_j( \vec r^\prime)$.  The convergence of this expansion is improved by adding an  homogeneous term and a one body contribution, thus satisfying the electron-electron and the electron-ion cusp conditions, respectively \cite{benzene}\cite{geminal}. The basis set used for the Jastrow includes $2s2p$ Gaussian orbitals.
By optimizing the coefficients $g_{i,j}$, we can treat in a non perturbative way the dynamical transitions  to high angular momentum atomic states for pairs of electrons localized around each atoms. As shown for the benzene \cite{benzene} and  water dimer \cite{water}, these transitions are responsible, at the first order of perturbation theory, for the weak attractive dispersive forces between atoms at large separation.

\begin{figure}
[ptb]
%\vskip 0.5cm
\begin{center}
\includegraphics[height=1.8in,width=2.9in]{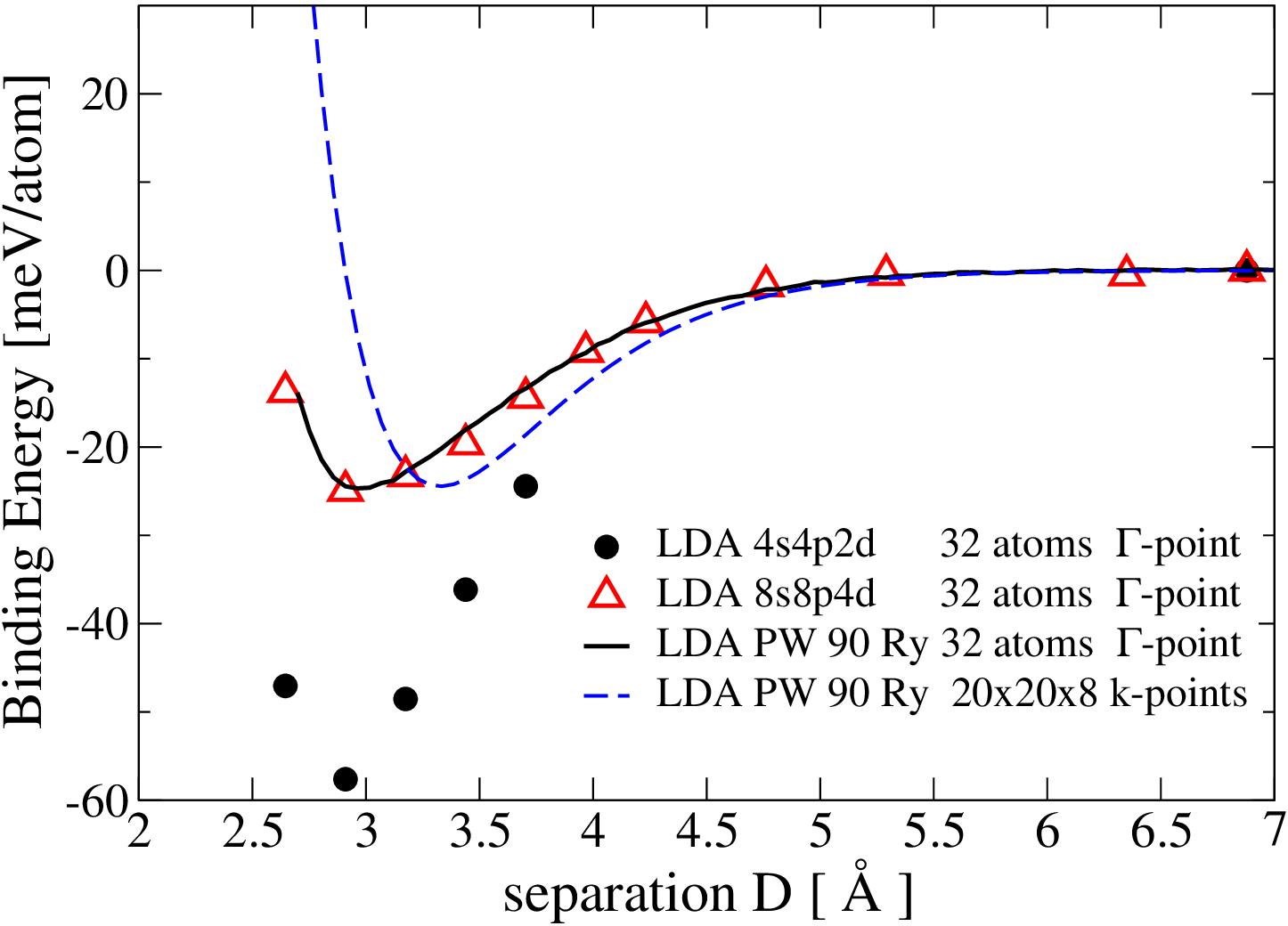}
\caption{ Binding Energy (BE) curve for graphite with AB stacking obtained at
the DFT-LDA level of theory, using plane waves (PW) and Gaussian basis sets. The results of PW calculations (solid black line) converged as a function of the kinetic energy cutoff ($90$ Ry) are in excellent agreement with those of $8s8p4d$ Gaussian  basis sets (red triangles), carried out with the same cell ($32$ atoms and the $\Gamma$ point). Smaller Gaussian basis sets ($4s4p2d$, full circle) yield a much larger BE and a slightly smaller equilibrium distance.  Fully converged PW calculations ($4$-atoms unit cell, $90$ Ry, $20\times20\times8 $ {\bf k}-points) yield
a BE of $24$ meV/atom and an equilibrium inter-layer position
D of $3.30$ \AA\ . Note that $D$ is significantly affected by convergence
as a function of {\bf k}-point sampling, while the value of the BE depends
weakly on it.}
\label{basis_set}
\end{center}
\end{figure}

In the following, the molecular orbitals $\psi_j$  in the Slater determinant   are obtained from a self-consistent DFT-LDA calculation. One may then optimize only the Jatrow factor, by keeping fixed the determinant built from LDA orbitals (hereafter referred to as J-DFT WF approach); alternatively one may simultaneously optimize both $J$ and the determinant  using the  method described in Ref.\cite{optimization}.  
 The minimal Gaussian basis set required to build an accurate determinant was chosen by comparing graphite binding energies (BE) \cite{endnote29} as obtained using plane waves (PW) and Gaussian basis sets (see Fig.\ref{basis_set}).  We note that at each atomic positions, PW calculations are free of basis set superposition errors and they can be converged by controlling one parameter, the kinetic energy cutoff. In the case of Gaussian, we used an even tempered local basis where the parameters $\alpha_l$ and $\beta_l$ of the Gaussian exponent $Z_{i} = \alpha_{l} \beta_{l}^{i}$ of each angular momentum $l$, were optimized by  performing a series of total energy LDA calculations.  We considered two basis sets: $4s4p2d$ and $8s8p4d$. For both of them 
%we optimized the parameters $\alpha_l$ and $\beta_l$ and 
we computed the binding energy (BE) of graphite at the LDA level for a system of $32$ atoms using only the $\Gamma$ point. The $8s8p4d$ basis set reproduces the same BE curve obtained with PW converged with respect to the kinetic energy cutoff ($90$Ry).
%% geometry

In our QMC calculations we simulated  a $2\times 2\times 1$ and $2\times 2\times 2$ super-cell with periodic boundary condition, containing $32$  ($128$ electrons) and $64$ atoms ($256$ electrons) respectively. The carbon valence-core interaction was described by a energy-consistent pseudopontential \cite{filippi}. In our calculations we fixed the in-plane geometry  to the one determined experimentally (C-C distance = $1.42$ \AA\ ). We verified, at the LDA level, that a change as large as $10\%$ in the in-plane carbon-carbon length affects the inter-plane equilibrium distance by less than $3\%$, while the BE changes by $1-2$meV/atom.

%% tabella energie 
In Table \ref{tabella1} we compare the results obtained by full wave function optimization with those of the  J-DFT WF approach for graphite at inter-layer distance $D=3.7$ \AA\ .
The latter provides a reasonably accurate variational guess. However, full optimization of the WF parameters (including the exponent of the basis set) yields a decrease of energy per atom of $34(3)$ meV/atom, which is of a relevant magnitude to the energy scale we wish to investigate in this work. 
%The LRDMC energy is less sensitive to the change in the nodal structure due to the simultaneous optimization of Jastrow factor and the determinant.
The LRDMC   energy is much less sensitive  to  the initial state (guiding function), used in this ground state projection technique.
In fact  the optimization of Jastrow and Determinant in the guiding function, leads to a consistent LRDMC energy.
  The quantitative agreement between the VMC and LRDMC calculations confirms that the key ingredients of the electron correlations are already included in our Jastrow factor.  In the following the results for the $2\times 2\times 1$ super-cell are obtained by fully optimizing the wavefunction, while, due to the computational cost of the optimiziation, we will present only LRDMC calculations for the $64$ atom system. 

\begin{small}
\begin{table}[t]%
\begin{tabular}
{cccc}\hline
Basis Set& Method& E (eV/atom) & $\sigma$(eV/atom)  \\
\hline
\hline
4s4p2d& Opt.  VMC   & $-154.428(3) $  &  $1.79(5)  $     \\
8s8p4d& J-DFT VMC   & $-154.505(1) $  &  $1.66(1) $   \\
8s8p4d& Opt. VMC   & $-154.540(1) $  &  $1.60(1)$   \\
\hline
4s4p2d& Opt. LRDMC &      $ -154.787(8) $       &  $\dots$        \\
8s8p4d& J-DFT    LRMDC & $ -154.891(5) $  &  $\dots$        \\
8s8p4d& Opt. LRDMC &      $ -154.899(5) $   &   $\dots$        \\
\hline
\end{tabular}
\caption{ Variational (VMC) and Diffusion (LRDMC) total energy and energy variance ($\sigma$) obtained using two different guiding wave functions for the $2\times2\times1$ super-cell at a separation distance $D=3.7$ \AA. $J-DFT$ denotes a wave function with optimized Jastrow and a determinant part from LDA calculations. "Opt." refers to the guiding wavefunction where both the Jastrow factor and the orbitals were simultaneously optimized.}
\label{tabella1}%
\end{table}
\end{small}

%size effects

After assessing the accuracy of the  guiding wave function, we considered finite-size (FS) effects. We expect the errors due to  FS  effects in the in-plane directions (i.e. $x$ and $y$ directions) to cancel out to a large extent \cite{drummond_needs}, as we compute energy {\it differences} between systems (graphite and graphene) with rather similar bonding and electronic properties. In-plane  FS errors arising from the kinetic and Hartree terms (one-body corrections)  can be treated within a standard DFT approach with appropriate ${\bf k}$-point sampling. %As  evidence that one-body corrections are well described by LDA, we observe that the 3B-DFT WF provides the same LRDMC energy as the fully optimized WF (Table \ref{tabella1}), indicating that the LDA orbitals are rather accurate for the graphite system.  
Other FS errors come from the artificial periodicity of the exchange-correlation hole due to the periodic Coulomb potential. Kwee, Zhang and Krakauer (KZK) ~\cite{KZK} proposed to calculate the two-body corrections within LDA where the exchange and correlation energy is replaced by the LDA energy parametrized for a finite system. We applied KZK corrections as implemented in the {\it PWSCF} code ~\cite{pwscf} \footnote{Corrections to the two-body term may also be treated using the method proposed in Ref.~\cite{chiesa},  which requires the knowledge of the structure factor $S({\bf q})$ for  ${\bf q} \to 0$ vectors. In principle $S({\bf  q})$ can be directly evaluated using QMC calculations. In practice, for semi-metallic graphite and graphene  small error bars are difficult to achieve. In Ref. \cite{sola} the FS correction schemes proposed in Ref. ~\cite{chiesa} and ~\cite{KZK} were shown to perform equally well.
}.

\begin{figure}
[ptb]
%\vskip 0.5cm
\begin{center}
\includegraphics[height=2.0in,width=2.9in]{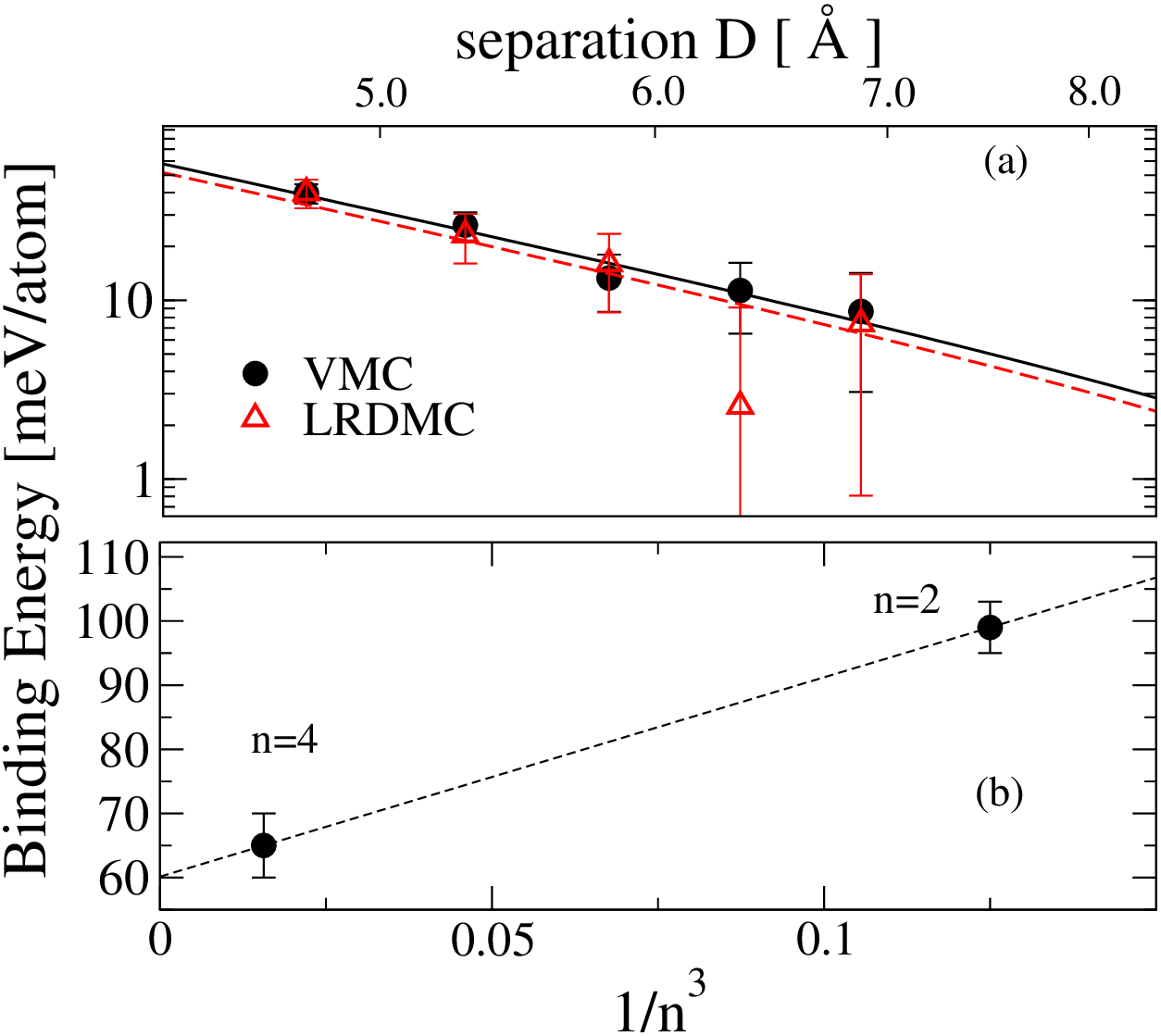}
\caption{Binding energy (BE) of graphite as a function of separation between planes (a), and of the number of layers ($n$) included in the periodic cell used in our calculations (b). Note the logarithmic scales. In (a) results obtained with VMC and LRDMC are reported. In (b) only LRDMC results are shown.
}
\label{total_log}
\end{center}
\end{figure}

We cannot rely on any error cancellation  in the $z$ direction, i.e. F.S. errors  due to a finite number of graphene layers in the simulation cells. In addition, the KZK method cannot  provide a robust correction scheme in this case due to the lack of long range effetcs  in the LDA exchange and correlation functionals. We estimate the long range behavior of the interaction between planes by fitting the results of calculations performed on the $2\times2\times1$ super-cell at distances $D>4$ \AA\ . These VMC (LRDMC) results are reported  in Fig. (\ref{total_log}) and show a  behavior $E(D) \sim D^{-\alpha}$ with $\alpha=4.2(1) (4.2(3))$. Although the LRDMC data are affected by larger error bars, we can safely conclude that $\alpha \geq 4$. We note that using the RPA applied to 2D systems, Dobson {\it et al.} \cite{dobson} found a power law behavior  $\sim D^{-3}\log(D/D_0)$ for infinite $\pi$-conjugated layers. One does not expect to see the asymptotic form of Ref. \cite{dobson}  in our work because the unusual interaction, arising from coupling between long-wavelength fluctuations in the plane,  it is expected to arise at much larger distances than those studied here \footnote{We note that sums of pair-wise forces proportional to $1/R^{6}$ yield  a behavior of the type $E \sim D^{-4}$. Of course our results showing a $\sim D^{-4}$ behavior are not to be taken as confirming that description of binding in graphite with pair-wise $1/R^6$ forces is correct.}
  
\begin{figure}
[ptb]
%\vskip 0.5cm
\begin{center}
\includegraphics[height=2.0in,width=2.7in]{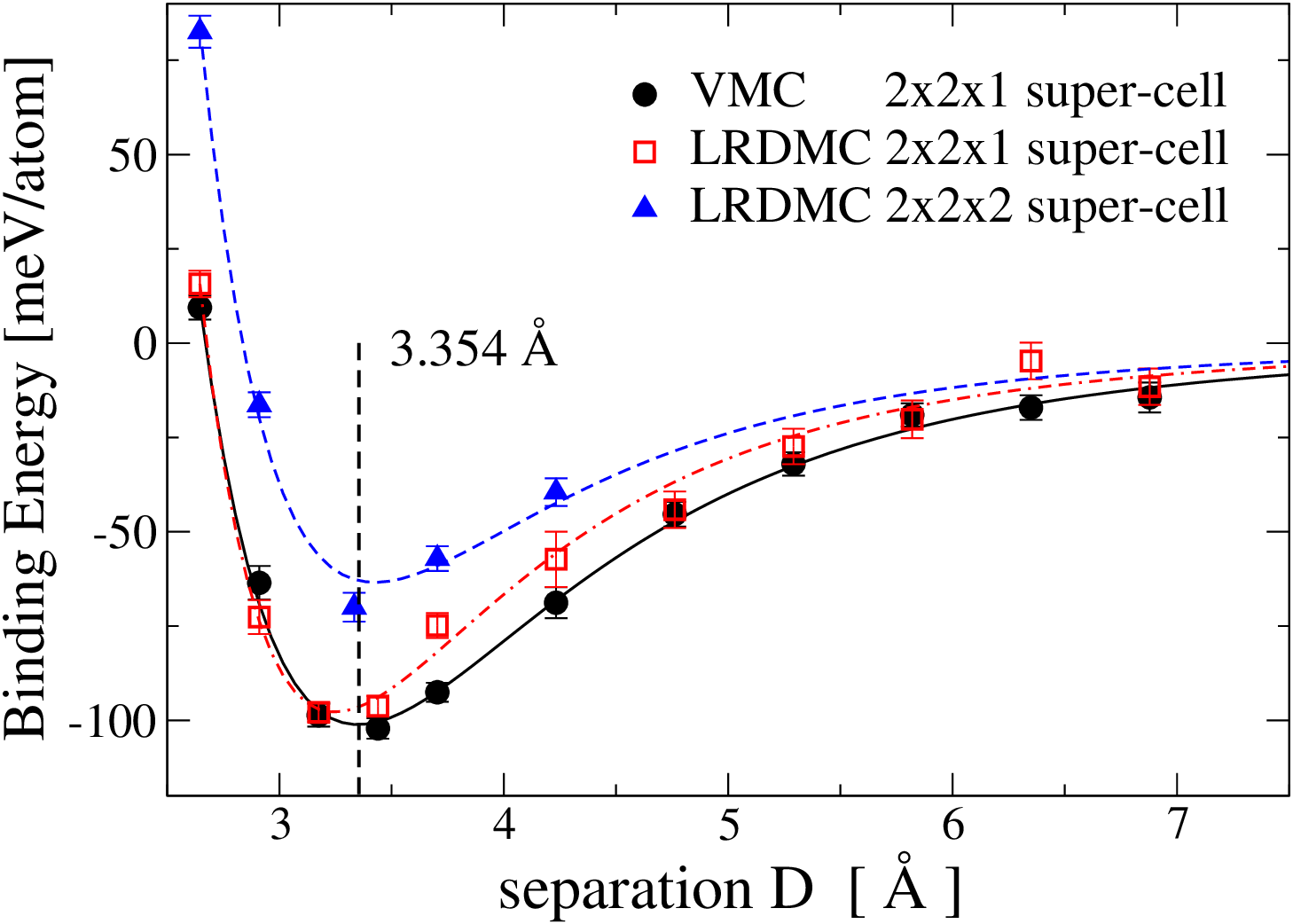}
\caption{ Binding Energy (BE) of graphite as obtained at the  VMC (black 
solid circles) and
LRDMC (red  squares) level of theory, using a  $2\times2\times1$ super-cell,  
and at the LRDMC level (blue
triangles)  with a $2\times2\times2$ super-cell. All data include finite size 
corrections using the scheme of Ref \cite{KZK}. Dotted and solid lines are 
obtained with a fit with the function $a \exp(\delta D) + b/D^4$. The experimental
inter-layer distance \cite{delhaes} is shown  by the dashed line. The large 
difference in BE between  the $2\times2\times1$  and $2\times2\times2$ super-cell calculations 
arise from large spurious interactions between periodic images of planes in the case of the $2\times2\times1$ cell. These spurious interactions are still present in the case of the $2\times2\times2$ cell, but they are greatly reduced, as shown by the difference ($\sim5$ meV/ atom) of BE obtained with 4 planes in  the cell and the extrapolated value.}
\label{vmc_lrdmc_final}
\end{center}
\end{figure}

Using the power law determined in our calculation we can  derive a scaling relation between the graphite BE and the number of layers $n$. Integrating over the super-cell volume we find that the total energy scales as $\sim 1/D^{3}_{max}$ where $D_{max}$ is the linear size of the super-cell in the $z$ direction, i.e as $\sim 1/n^3$. In Fig. (\ref{total_log}b) we report the BE  obtained within the LRDMC method as a function of $\sim 1/n^3$. The BE curves close to the minimum for the $32$ and $64$ atoms cells are reported in Fig. (\ref{vmc_lrdmc_final}). Extrapolating the results reported in Fig. (\ref{total_log}b) to an infinite number of layers, we obtain a value of the BE of $60 (5)$ meV/atom. This is reduced to $56(5)$, after adding zero point motion ($\sim 2$ meV/atom) and lattice vibrational contributions at 300 K ($\sim 2$ meV/atom), as computed from vibrational free energies using the data of Ref. \cite{marzari} for phonon frequencies.

Absorption experiments of aromatic molecules on graphite ~\cite{zacharia} provide a measurement of the  EE, while the cleavage energy is estimated to be $18\%$ larger than exfoliation, on the basis of force field calculations ~\cite{girifalco}. The cleavage energy is close, althogh not identical ot the BE defined here \cite{endnote29}. Therefore our comparison with experiment can only be indirect, as we computed BE, while experiment reports EE. 
Nevertheless it appears that our computed BE ($\sim 56$ meV/atom )  is in good agreement with the  estimate for the cleavage energy from the most recent experiments: $62$ meV/atom. We note that in the analysis of experimental results, one makes use of fitted force fields to evaluate the contribution to EE of carbon-hydrogen bonds. This is needed because an extrapolation is made on hydrocarbon adsorption energies, as a function of the number of C atoms. Although the force field parameters were adjusted to experimental data, it is unclear whether the assumption of additivity of forces close to the minimum is fully justified. In addition, the ratio between EE and BE is at present unknown and could only be estimated.
 
While the value of the BE can be extrapolated for an infinite number of planes (as in Fig. (\ref{total_log})), the value of the equilibrium separation cannot. From the minima of the curves reported in Fig. (\ref{vmc_lrdmc_final})  we obtain  $3.350(24)$ and $3.243(26)$ \AA\ at the VMC and LRDMC level, respectively, for the $2\times2\times1$ cell, and $3.426(36)$ \AA\ at the LRDMC level, for the $2\times2\times2$ cell. Difficulties arising from very flat BE curves and, most importantly, from the lack of an extrapolation procedure as a function of the number of layers prevent us to find a fully converged equilibrium bond-length. The value found  for the $2\times2\times2$  cell is in good agreement with experiment ($2\%$ overestimate) \footnote{ For the $2\times2\times2$ super-cell the position of the minimum was estimated considering LRDMC calculations at lattice space of $a=0.4$ \AA\ . In the case of the $2\times2\times1$ cell we observed that the extrapolation procedure for $a \to 0 $ does not affect the equilibrium position.}.

In conclusion, we have investigated the bonding properties of graphite in AB stacking, providing for the first
time an estimate of the binding energy and long range behavior of the total energy based on {\it ab-initio}, many body theory. Our calculated binding energy is in good agreement with most recent experiments, providing  a benchmark result for future calculations and further experimental measurements. The interaction energy between planes varies as $D^{-4.2}$ , i.e with a power law close to that found for two semiconducting planes.

 We acknowledge usefull discussions and correspondance with D.Lu, C.Attaccalite, D.C.Langreth, T.Hertel, M. Casula, D. Ceperly.
We thank N.Marzari and N. Bonini for discussions and for providing the data for the phonons frequences and  J.F. Dobson for a critical reading of the manuscript. L.S. and G.G. acknowledge the support by  DOE  DE-FC02-06ER25794. S.S. thanks CINECA and MIUR PRIN'07.

%\bibliography{bibliografia}

\begin{thebibliography}{26}
\expandafter\ifx\csname natexlab\endcsname\relax\def\natexlab#1{#1}\fi
\expandafter\ifx\csname bibnamefont\endcsname\relax
  \def\bibnamefont#1{#1}\fi
\expandafter\ifx\csname bibfnamefont\endcsname\relax
  \def\bibfnamefont#1{#1}\fi
\expandafter\ifx\csname citenamefont\endcsname\relax
  \def\citenamefont#1{#1}\fi
\expandafter\ifx\csname url\endcsname\relax
  \def\url#1{\texttt{#1}}\fi
\expandafter\ifx\csname urlprefix\endcsname\relax\def\urlprefix{URL }\fi
\providecommand{\bibinfo}[2]{#2}
\providecommand{\eprint}[2][]{\url{#2}}

\bibitem[{\citenamefont{Castro~Neto}(2009)}]{castro}
\bibinfo{author}{\bibfnamefont{A.H.}} \bibnamefont{Castro~Neto}
\bibinfo{author}{\bibnamefont{{\it et al.}}},
  \bibinfo{journal}{Rev. Mod. Phys.} \textbf{\bibinfo{volume}{81}},
  \bibinfo{pages}{109} (\bibinfo{year}{2009}).

\bibitem[{\citenamefont{Deshpande}(2009)}]{gg2}
\bibinfo{author}{\bibfnamefont{A.} \bibnamefont{Deshpande}}
 \bibinfo{author}{\bibnamefont{{\it et al.}}},
  \bibinfo{journal}{Phys. Rev. B} \textbf{\bibinfo{volume}{79}},
  \bibinfo{pages}{205411} (\bibinfo{year}{2009}).

\bibitem[{\citenamefont{Dobson }(2006)}]{dobson}
\bibinfo{author}{\bibfnamefont{J.F.}~\bibnamefont{Dobson}},
\bibinfo{author}{\bibfnamefont{A.}~\bibnamefont{White}} \bibnamefont{and}
\bibinfo{author}{\bibfnamefont{A.}~\bibnamefont{Rubio}}, 
  \bibinfo{journal}{{Phys. Rev.  Lett.}} \textbf{\bibinfo{volume}{96}}, \bibinfo{pages}{073201}
  (\bibinfo{year}{2006}).

\bibitem[{\citenamefont{Gould {\it et al.}}(2009)}]{DobsonPRB2009}
\bibinfo{author}{\bibfnamefont{T.}~\bibnamefont{Gould}},  
\bibinfo{author}{\bibfnamefont{E.}~\bibnamefont{Gray}} \bibnamefont{and}
\bibinfo{author}{\bibfnamefont{J.F.}~\bibnamefont{Dobson}}, 
\bibinfo{journal}{Phys. Rev. B}
  \textbf{\bibinfo{volume}{79}}, \bibinfo{pages}{113402}
  (\bibinfo{year}{2009}).

\bibitem[{\citenamefont{Kohn and Sham}(1965)}]{KS}
\bibinfo{author}{\bibfnamefont{W.}~\bibnamefont{Kohn}} \bibnamefont{and}
  \bibinfo{author}{\bibfnamefont{L.}~\bibnamefont{Sham}},
  \bibinfo{journal}{Phys. Rev.} \textbf{\bibinfo{volume}{140}},
  \bibinfo{pages}{1133} (\bibinfo{year}{1965}).

\bibitem[{\citenamefont{Perdew }(1996)\citenamefont{Perdew, Burke, and
  Ernzerhof}}]{gga}
\bibinfo{author}{\bibfnamefont{J.}~\bibnamefont{Perdew}},
\bibinfo{author}{\bibfnamefont{K.}~\bibnamefont{Burke}}  \bibnamefont{and}
\bibinfo{author}{\bibfnamefont{M.}~\bibnamefont{Ernzerhof}} 
  \bibinfo{journal}{Phys. Rev. Lett.} \textbf{\bibinfo{volume}{77}},
  \bibinfo{pages}{3865} (\bibinfo{year}{1996}).

\bibitem[{\citenamefont{Chakarova-Kack }(2006)}]{langreth}
\bibinfo{author}{\bibfnamefont{S.}~\bibnamefont{Chakarova-Kack}}
\bibinfo{author}{\bibnamefont{{\it et al.}}},
  \bibinfo{journal}{Phys. Rev. Lett.} \textbf{\bibinfo{volume}{96}},
  \bibinfo{pages}{146107} (\bibinfo{year}{2006}).

\bibitem[{\citenamefont{Hasegawa and Nishidate}(2004)}]{empirical}
\bibinfo{author}{\bibfnamefont{M.}~\bibnamefont{Hasegawa}} \bibnamefont{and}
  \bibinfo{author}{\bibfnamefont{K.}~\bibnamefont{Nishidate}},
  \bibinfo{journal}{{Phys. Rev. B}} \textbf{\bibinfo{volume}{70}},
  \bibinfo{pages}{205431} (\bibinfo{year}{2004}).

\bibitem[{\citenamefont{Grimme and {\it et al.}}({2007})}]{grimme}
\bibinfo{author}{\bibfnamefont{S.}~\bibnamefont{Grimme}}, 
\bibinfo{author}{\bibfnamefont{C.}~\bibnamefont{Muck-Lichtenfeld }} \bibnamefont{and}
\bibinfo{author}{\bibfnamefont{J.}~\bibnamefont{Antony}}
 \bibinfo{journal}{{Jour. Phys.  Chem. C}} \textbf{\bibinfo{volume}{{111}}}, \bibinfo{pages}{11199}
  (\bibinfo{year}{{2007}}).

\bibitem[{\citenamefont{Zacharia and {\it et al.}}(2004)}]{zacharia}
\bibinfo{author}{\bibfnamefont{R.}~\bibnamefont{Zacharia}}, 
\bibinfo{author}{\bibfnamefont{H.}~\bibnamefont{Ulbricht}} \bibnamefont{and}
\bibinfo{author}{\bibfnamefont{T.}~\bibnamefont{Hertel}}
 \bibinfo{journal}{{Phys. Rev. B}} \textbf{\bibinfo{volume}{69}}, \bibinfo{pages}{155406}
  (\bibinfo{year}{2004}).

\bibitem[{\citenamefont{Benedict and {\it et al.}}(1998)}]{benedict}
\bibinfo{author}{\bibfnamefont{L.}~\bibnamefont{Benedict}} 
  \bibinfo{author}{\bibnamefont{{\it et al.}}}, \bibinfo{journal}{Chem. Phys.
  Lett.} \textbf{\bibinfo{volume}{286}}, \bibinfo{pages}{490}
  (\bibinfo{year}{1998}).

\bibitem[{\citenamefont{Girifalco and Lad}(1956)}]{girifalco}
\bibinfo{author}{\bibfnamefont{L.}~\bibnamefont{Girifalco}} \bibnamefont{and}
  \bibinfo{author}{\bibfnamefont{R.}~\bibnamefont{Lad}},
  \bibinfo{journal}{Jour. Chem. Phys.} \textbf{\bibinfo{volume}{25}},
  \bibinfo{pages}{693} (\bibinfo{year}{1956}).

\bibitem[{\citenamefont{Foulkes }(2001)\citenamefont{Foulkes, Mitas,
  Needs, and Rajagopal}}]{review}
\bibinfo{author}{\bibfnamefont{W.}~\bibnamefont{Foulkes}}
\bibinfo{author}{\bibnamefont{{\it et al.}}},
  \bibinfo{journal}{Rev. Mod. Phys.} \textbf{\bibinfo{volume}{73}},
  \bibinfo{pages}{33} (\bibinfo{year}{2001}).

\bibitem[{\citenamefont{Drummond and Needs}(2007)}]{drummond_needs}
\bibinfo{author}{\bibfnamefont{N.~D.} \bibnamefont{Drummond}} \bibnamefont{and}
  \bibinfo{author}{\bibfnamefont{R.~J.} \bibnamefont{Needs}},
  \bibinfo{journal}{Phys. Rev. Lett.} \textbf{\bibinfo{volume}{99}},
  \bibinfo{pages}{166401} (\bibinfo{year}{2007}).

\bibitem[{\citenamefont{Sorella  {\it et al.}}(2007)}]{benzene}
\bibinfo{author}{\bibfnamefont{S.}~\bibnamefont{Sorella}},
\bibinfo{author}{\bibfnamefont{M.}~\bibnamefont{Casula}} \bibnamefont{and}
\bibinfo{author}{\bibfnamefont{D.}~\bibnamefont{Rocca}}
\bibinfo{journal}{Jour. Chem.   Phys.} \textbf{\bibinfo{volume}{127}}, \bibinfo{pages}{014105}
  (\bibinfo{year}{2007}).

\bibitem[{\citenamefont{Casula and {\it et al.}}(2005)}]{casula_lrdmc}
\bibinfo{author}{\bibfnamefont{M.}~\bibnamefont{Casula}},
\bibinfo{author}{\bibfnamefont{C.}~\bibnamefont{Filippi}}   \bibnamefont{and}
\bibinfo{author}{\bibfnamefont{S.}~\bibnamefont{Sorella}}
\bibinfo{journal}{Phys. Rev. Lett.} \textbf{\bibinfo{volume}{95}}, \bibinfo{pages}{100201}
  (\bibinfo{year}{2005}).

\bibitem[{\citenamefont{Sorella}()}]{TRVB}
\bibinfo{author}{\bibfnamefont{S.}~\bibnamefont{Sorella}},
  \bibinfo{journal}{https://qe-forge.org/projects/turborvb/ }.

\bibitem[{\citenamefont{Casula {\it et al.}}({2004})}]{geminal}
\bibinfo{author}{\bibfnamefont{M.}~\bibnamefont{Casula}},
\bibinfo{author}{\bibfnamefont{C.}~\bibnamefont{Attacalite}} \bibnamefont{and}
\bibinfo{author}{\bibfnamefont{S.}~\bibnamefont{Sorella}}
\bibinfo{journal}{{Jour. Chem.  Phys.}} \textbf{\bibinfo{volume}{{121}}}, 
\bibinfo{pages}{{7110}}   (\bibinfo{year}{{2004}}).

\bibitem[{\citenamefont{Sterpone }(2008)\citenamefont{Sterpone}}]{water}
\bibinfo{author}{\bibfnamefont{F.}~\bibnamefont{Sterpone}}
\bibinfo{author}{\bibnamefont{{\it et al.}}},
  \bibinfo{journal}{Jour. Chem. Theory Comp.} \textbf{\bibinfo{volume}{4}},
  \bibinfo{pages}{1428} (\bibinfo{year}{2008}).

\bibitem[{\citenamefont{Umrigar}(2007)}]{optimization}
\bibinfo{author}{\bibfnamefont{C.J.}~\bibnamefont{Umrigar}}
\bibinfo{author}{\bibnamefont{{\it et al.}}},
  \bibinfo{journal}{Phys. Rev. Lett.} \textbf{\bibinfo{volume}{98}},
  \bibinfo{pages}{110201} (\bibinfo{year}{2007}).

\bibitem[{\citenamefont{Burkatzki.}(2007)\citenamefont{Burkatzki,
  Filippi, and Dolg}}]{filippi}
\bibinfo{author}{\bibfnamefont{M.}~\bibnamefont{Burkatzki}},
 \bibinfo{author}{\bibfnamefont{C.}~\bibnamefont{Filippi}}   \bibnamefont{and}
 \bibinfo{author}{\bibfnamefont{M.}~\bibnamefont{Dolg}} 
  \bibinfo{journal}{Jour. Chem. Phys.} \textbf{\bibinfo{volume}{126}},
  \bibinfo{pages}{234105} (\bibinfo{year}{2007}).

\bibitem[{\citenamefont{Charlier }(1994)\citenamefont{Charlier, Gonze,
  and Michenaud}}]{charlier}
\bibinfo{author}{\bibfnamefont{J.C.}~\bibnamefont{Charlier}}, 
\bibinfo{author}{\bibfnamefont{X.}~\bibnamefont{Gonze}}    \bibnamefont{and}
\bibinfo{author}{\bibfnamefont{J.P.}~\bibnamefont{Michenaud}}
  \bibinfo{journal}{Europhys. Lett.} \textbf{\bibinfo{volume}{28}},
  \bibinfo{pages}{403} (\bibinfo{year}{1994}).

\bibitem[{\citenamefont{Kwee and {\it et al.}}({2008})}]{KZK}
\bibinfo{author}{\bibfnamefont{H.}~\bibnamefont{Kwee}}, 
\bibinfo{author}{\bibfnamefont{S.W.}~\bibnamefont{Zhang}} \bibnamefont{and}
\bibinfo{author}{\bibfnamefont{H.}~\bibnamefont{Krakauer}}
 \bibinfo{journal}{{Phys. Rev.  Lett.}} \textbf{\bibinfo{volume}{{100}}}, \bibinfo{pages}{126404}
  (\bibinfo{year}{{2008}}).

\bibitem[{\citenamefont{Baroni}()}]{pwscf}
\bibinfo{author}{\bibfnamefont{S.}~\bibnamefont{Baroni}}
\bibinfo{author}{\bibnamefont{{\it et al.}}},
  \bibinfo{journal}{http://www.pwscf.org}.

\bibitem[{\citenamefont{Chiesa  {\it et al.}}(2006)}]{chiesa}
\bibinfo{author}{\bibfnamefont{S.}~\bibnamefont{Chiesa}} 
\bibinfo{author}{\bibnamefont{{\it et al.}}}, 
\bibinfo{journal}{Phys. Rev.  Lett.} 
\textbf{\bibinfo{volume}{97}}, \bibinfo{pages}{076404} (\bibinfo{year}{2006}).

\bibitem[{\citenamefont{Sola}(2009)}]{sola}
\bibinfo{author}{\bibfnamefont{E.}~\bibnamefont{Sola}},
\bibinfo{author}{\bibfnamefont{J.P.}~\bibnamefont{Brodholt}} \bibnamefont{and}
\bibinfo{author}{\bibfnamefont{D.}~\bibnamefont{Alf\'e}},
  \bibinfo{journal}{Phys. Rev. B} \textbf{\bibinfo{volume}{79}},
  \bibinfo{pages}{024107} (\bibinfo{year}{2009}).

\bibitem[{\citenamefont{Delha\'es}()}]{delhaes}
\bibinfo{author}{\bibfnamefont{P.}~\bibnamefont{Delha\'es}},
  \bibinfo{journal}{Graphite and Precursors, Gordon and Breach Science Publishers},
(\bibinfo{year}{2001}).



\bibitem[{\citenamefont{Mounet and Marzari}(2005)}]{marzari}
\bibinfo{author}{\bibfnamefont{N.}~\bibnamefont{Mounet}} \bibnamefont{and}
  \bibinfo{author}{\bibfnamefont{N.}~\bibnamefont{Marzari}},
  \bibinfo{journal}{Phys. Rev. B} \textbf{\bibinfo{volume}{71}},
  \bibinfo{pages}{205214} (\bibinfo{year}{2005}).

\end{thebibliography}

\end{document}